\documentclass[preprint]{ptephy_v1}

\preprintnumber{OCU-PHYS-469, AP-GR-141} 

\newcommand{\be}{\begin{equation}}
\newcommand{\ee}{\end{equation}}
\newcommand{\ba}{\begin{eqnarray}}
\newcommand{\ea}{\end{eqnarray}}
\newcommand{\ban}{\begin{eqnarray*}}
\newcommand{\ean}{\end{eqnarray*}}

\newcommand{\blue}[1]{{{\textcolor{blue}{#1}}}}

\newcommand{\cE}{\ensuremath{{\cal E}}}
\newcommand{\Gm}{\ensuremath{\varGamma}}

\newcommand{\veps}{\varepsilon}

\newcommand{\tm}{\tilde{m}}
\newcommand{\tq}{\tilde{q}}

\newcommand{\rc}{r_{\rm c}}

\begin{document}

\title{Non-linear collisional Penrose process \\[.5em]
{\it -- How large energy can a black hole release? --}}


\author{Ken-ichi Nakao}
\affil{Department of Mathematics and Physics, Graduate School of Science, Osaka City University, Osaka 558-8585, Japan \email{knakao-at-sci.osaka-cu.ac.jp}}

\author{Hirotada Okawa}
\affil{Yukawa Institute for Theoretical Physics, Kyoto University, Kyoto 606-8502, Japan}
\affil{Advanced Research Institute for Science and Engineering, Waseda University, Tokyo 169-8555, Japan}


\author{Kei-ichi Maeda}
\affil{Department of Physics, Waseda University, Tokyo 169-8555, Japan}


\begin{abstract}%
Energy extraction from a rotating or charged black hole is one of fascinating issues in general relativity. 
The collisional Penrose process is one of such extraction mechanisms and   
has been reconsidered intensively since Ba\~nados, Silk and West pointed out the physical importance of 
very high energy collisions around a maximally rotating black hole. In order to get results analytically,  
the test particle approximation has been adopted so far. 
Successive works based on this approximation scheme have not yet revealed the upper bound 
on the efficiency of the energy extraction because of lack of the back reaction. 
In the Reissner-Nordstr\"{o}m spacetime, by fully taking into account the self-gravity of the shells, 
we find that there is an upper bound on the extracted energy, which is consistent with the area law of a black hole. 
We also show one particular scenario in which the almost maximum energy extraction is achieved 
even without the Ba\~nados-Silk-West collision.    

\end{abstract}

\subjectindex{E01, E31}

\maketitle

\section{Introduction}\label{sec:intro}

Energy extraction from a black hole is one of the 
interesting and important issues not only in  general relativity but also in  astrophysics 
(engines of $\gamma$-ray bursts, energy sources of jets from AGN, origins of ultra-high-energy cosmic rays, etc.).
In 1969~\cite{Penrose1969}, Penrose pointed out that it is possible to extract the
rotational energy of a Kerr black hole, which is a stationary and axi-symmetric rotating black hole,  
through the decay of a particle falling from the infinity to create two particles 
in the ergo-region, in the case that one is bounded with negative energy, whereas 
the other escapes to infinity with positive energy. 
Successive works revealed that this mechanism does not work quite efficiently 
in the astrophysical situation~\cite{Bardeen1972,Wald1974}. 
A bit modified version of the Penrose process called the 
collisional Penrose process, in which two particles collide with each other in the ergo-region instead
of a single particle decay, was first noticed by Piran, Shaham and Katz~\cite{Piran1975},
but its efficiency as modest as the original process was reported.

Recently, the collisional Penrose process again attracts people 
since Ba\~nados, Silk and West (BSW) showed 
that there is no upper bound on the center-of-mass energy of two particles colliding with each other 
almost at the event horizon of an extremal Kerr black hole~\cite{Banados2009}. 
This fact does not necessarily mean the unbounded energy extraction from the black hole,
as the particle escaping to the infinity wastes its energy to run up the deep gravitational potential. 
Nevertheless, some works consistently show that 
fine-tuned parameters of the particles result in the energy output about 14 times larger than the 
input energy~\cite{Bejger2012,Harada2012,Schnittman2014,Ogasawara2015,Leiderschneider2016,Zaslavskii2016}.
It is worthwhile to notice that the same conclusion is 
derived even though the deformation of the event horizon caused by energetic particles swallowed by  
the black hole is taken into account in accordance with the hoop conjecture~\cite{Hod2016}.
More efficient extraction mechanism of the energy from a black hole, 
which has been named the super-Penrose process, 
was suggested \cite{Berti2014,Patil2015}, but there is still an argument \cite{Leiderschneider2016}.

In order to know how large energy a black hole can really release through the Penrose process, 
one should fully take into account the nonlinearity of the Einstein equations.
It is much complicated and not so easy 
to treat a Kerr black hole with the gravitational backreaction by the particles. 
Here it is worthwhile to notice that the similar phenomenon to 
the BSW collision~\cite{Zaslavskii2010} and the collisional Penrose process ~\cite{Zaslavskii2012,Nemoto2013} 
can occur in the case of the Reissner-Nordstr\"{o}m black hole which is a spherically symmetric charged 
black hole. 
The BSW collision can also occur between two infinitesimally thin charged 
dust shell concentric to the Reissnr--Nordstr\"{o}m black hole although 
non-linear effects are taken into account through Israel's formalism~\cite{Kimura2011}. 
In this paper, we shall study the collisional Penrose process in the similar situation to that studied in Ref.~\cite{Kimura2011}
 and analyze the energy extraction efficiency. 

This paper is organized as follows.
In Section~\ref{sec:situation}, we explain our setup and derive the equations of motion for a 
spherically symmetric infinitesimally thin charged shell 
in accordance with Israel's formalism. Also in this section, we derive the formulation to get the  
conditions of two thin shells concentric with each other just after a collision with the mass transfer 
by imposing the 4-momentum conservation.   
We estimate the maximum extraction from the central black hole by analytic means in Section~\ref{sec:analytic}.
Section~\ref{sec:conclusion} is devoted to concluding our analyses.

We adopt the abstract index notation: small Latin indices indicate a type of a tensor, 
whereas Greek indices denote components of a tensor with respect to the coordinate basis vector~\cite{Wald}. 
We also adopt the signature of the metric and the convention of the Riemann tensor used in Ref.~\cite{Wald}. 
The geometrized unit is adopted. 

\section{Setup and basic equations}\label{sec:situation}
\subsection{Setup}
We consider two spherically symmetric shells concentric with each other. 
Each shell is infinitesimally thin and generates a timelike hypersurface
through its motion. We will often refer this hypersurface as a shell. 
These shells will collide with each other, and 
divide the spacetime into four regions (see Fig.1). 
Before the collision, we call these shells Shell 1 and Shell 2, respectively. 
After the collision, the shell which faces on a region together with  
Shell 2 is called Shell 3, and the other shell is called Shell 4. 
The region whose boundary is formed by Shell 1 and Shell 4 
is called Region 1,  
while the region between Shell 1 and Shell 2 is called Region 2. 
Similarly, the region whose boundary is formed by Shell 2 and Shell 3 
is called Region 3, and the region 
between Shell 3 and Shell 4 is called Region 4.
For notational convenience, Region 1 is often called Region 5. 
Hereafter, we use capital Latin indices, $I$, $J$ and $K$, 
to specify a shell or a region: $I$ runs from $1$ to $4$,
$J$ takes the values $1$ and $2$, which represents the shells
before the collision, and $K=3$ and $4$, 
labeling the shells after the collision. 

\begin{figure}[!h]
\centering\includegraphics[width=2.5in]{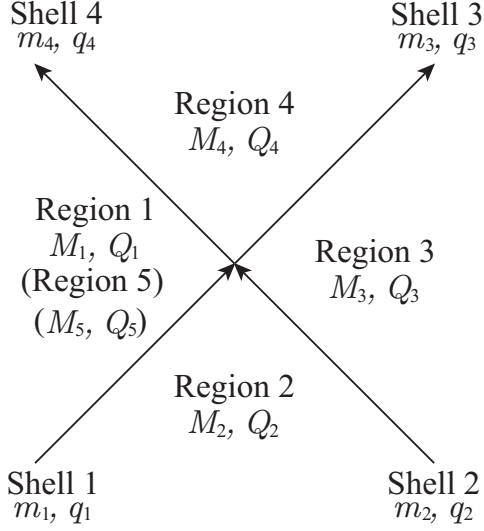}
\caption{ A schematic diagram of the collision between two spherical shells. 
 The vertical direction is temporal, while the horizontal direction is radial.
 }
\label{initial}
\end{figure}

\subsection{Equations of motion for shells}

Let $n_{I}^{a}$ be a unit outward space-like vector normal to Shell $I$, and 
define the projection operator as $h^{a}_{Ib} \equiv
\delta^{a}_{b} - n_{I}^{a} n_{Ib}$. 
Each shell is characterized by the surface stress-energy tensor which is given by
$$
 S^{ab}_{I} \equiv \lim_{\epsilon\rightarrow 0} 
  \int_{-\epsilon}^{+\epsilon}
  h^{a}_{Ic}h^{b}_{Id}T^{cd}_{I}\ dz,
$$
where $z$ is a Gaussian normal coordinate ($z=0$ on the shell). 

The extrinsic curvature of a timelike hypersurface 
generated by the motion of Shell $I$ is defined by
$$
K_{Iab} \equiv -h_{Ia}^ch_{Ib}^d\nabla_c n_{Id},
$$
where $\nabla_a$ is the covariant derivative. 

The Einstein equations lead to the jump condition for the extrinsic curvatures  
and the conservation law for $S_{I}^{ab}$\cite{Israel}: 
\begin{align}
  K_{Iab}|_{+} - K_{Iab}|_{-} &=  8\pi
 \left( S_{Iab} 
 -{1\over 2}h_{Iab} S^{c}_{Ic}\right), \label{eq:jump1}\\
  S_{I}^{ab} \left( K_{Iab}|_{+} +K_{Iab}|_{-} 
    \right) &= 0, \quad \label{eq:jump2}
\end{align}
and
\begin{equation}
D_{Ib}S_{I}^{ab}=0,\label{eq:jump3}
\end{equation}
where the quantity with the subscript $+$ is defined in the region to which  
the unit normal $n_I^a$ points, whereas that with the subscript $-$ is  
evaluated on another side.

Now let us turn the spherically symmetric case. 
We assume that the line element in Region $I$ is given in the form
\begin{equation}
  ds_{I}^2 = - f_{I}(r) dt^2 
  + {dr^2 \over f_{I}(r)} 
  + r^2 (d\theta^{2}+\sin^{2}\theta d\varphi^{2}), \label{eq:st-line}
\end{equation}
where $f_I(r)$ is not specified in this section so that the results obtained here is applicable 
to various cases: a vacuum spacetime, one with the Maxwell field, 
one  with a cosmological constant, and so on.  

The components of the 4-velocity of Shell $I$ are expressed as
\begin{equation}
u^{\alpha}_{I}|_{\pm}=\left(\dot{t}_{I\pm},
  \dot{r}_{I},0,0\right), \label{eq:u-def}
\end{equation}
where an over dot represents a derivative with respect to the proper time naturally defined on the shell.
Here note that the time coordinate, $t$, is not continuous across 
the shell, although the circumferential radius, $r$, the azimuthal angle, $\theta$, and the polar 
angle, $\varphi$, are everywhere continuous. 
Hence two different time coordinates ${t}_{I\pm}$ are assigned to 
each shell, and there are two kinds of time components of the
4-velocity.  
Using these components of the 4-velocity,
we obtain the components of the unit vector normal to Shell $I$ as
\begin{equation}
n_{I\alpha}|_{\pm}=\left(-\dot{r}_{I},\dot{t}_{I\pm},0,0\right). \label{eq:n-def}
\end{equation}

The surface-stress-energy tensor of the spherical shell takes the following form
$$
S_{I}^{ab}=\sigma_{I}u_{I}^{a}u_{I}^{b}+{\cal P}_{I}H^{ab}_{I},
$$
where $\sigma_{I}$ is the surface energy density, 
${\cal P}_{I}$ corresponds to the tangential pressure, 
and $H^{ab}_{I}\equiv h^{ab}_{I}+u^{a}_{I}u^{b}_{I}$ 
is the 2-sphere metric with the radius $r_{I}$. 
Then the conservation law (\ref{eq:jump3}) leads to 
\begin{equation}
\dot{m}_{I}=-8\pi{\cal P}_{I}r_{I}\dot{r}_{I},
\label{eq:conservation}
\end{equation}
where 
\begin{equation}
m_{I}\equiv 4\pi  \sigma_{I}r_{I}^{2}\label{eq:pmass-def}
\end{equation}
is the proper mass of Shell $I$. 
In the case of ${\cal P}_{I}=0$, we often call Shell $I$ a dust shell and 
Eq.~(\ref{eq:conservation}) implies that $m_{I}$ is constant. 
On the other hand, in the case of non-vanishing ${\cal P}_{I}$, 
$m_{I}$ depends on the proper time,  if the shell $I$ is moving. 
We assume the reasonable energy conditions, so that $m_{I}\geq 0$. 

Now, we assume  the outward normal $n_{J}^a$ which is directed from region $J$ to region $J+1$, whereas 
the direction of $n_{K}^a$ is from region $K+1$ to region $K$. 
This assumption implies, together with Eq.~(\ref{eq:n-def}),  
that the circumferential radius $r$ is increasing across the shell $J$ (shell $K$)  
from region $J$ (region $K+1$) to region $J+1$ (region $K$).  
Then, the junction condition (\ref{eq:jump1}) leads to
\begin{align}
 \dot{r}_{J}^{2} 
 &=\left(\cE_J-\frac{m_J}{2r_J}\right)^2-f_{J+1}(r_J)
 =\left(\cE_J+\frac{m_J}{2r_J}\right)^2-f_{J}(r_J), \label{eq:motion-J}
\end{align}
and 
\begin{align}
 \dot{r}_{K}^{2} 
 &=\left(\cE_K-\frac{m_K}{2r_K}\right)^2-f_{K}(r_K)
 =\left(\cE_K+\frac{m_K}{2r_K}\right)^2-f_{K+1}(r_K), \label{eq:motion-K}
\end{align}
where 
\begin{align}
 \cE_{J}
&\equiv  \frac{r_{J}}{2m_J}\left[f_{J}(r_J)-f_{J+1}(r_J)\right], \label{eq:energy-J} \\
 \cE_{K}
&\equiv  \frac{r_{K}}{2m_K}\left[f_{K+1}(r_K)-f_{K}(r_K)\right]
\,.
\label{eq:energy-K} 
\end{align}
As shown later, $\cE_I$ corresponds to the specific Misner-Sharp (MS) energy \cite{Misner1964} 
(MS energies per unit mass) of Shell $I$.

From the normalization of 4-velocity 
and Eqs.~(\ref{eq:motion-J}) and (\ref{eq:motion-K}), we obtain 
\begin{eqnarray}
 \dot{t}_{J+}
&=&{1\over f_{J+1}(r_{J})}
 \left(\cE_{J}-{m_{J}\over 2r_{J}}\right), \label{eq:dt-def1p}\\
 \dot{t}_{J-}
&=&{1\over f_{J}(r_{J})}
 \left(\cE_{J}+{m_{J}\over2r_{J}}\right),\label{eq:dt-def1m}
\end{eqnarray}
and
\begin{eqnarray}
 \dot{t}_{K+}
&=&{1\over f_{K}(r_{K})}
 \left(\cE_{K}-{m_{K}\over2r_{K}}\right), \label{eq:dt-def2p}\\
 \dot{t}_{K-}
&=&{1\over f_{K+1}(r_{K})}
 \left(\cE_{K}+{m_{K}\over2r_{K}}\right).\label{eq:dt-def2m}
\end{eqnarray}

\subsection{Momentum conservation}

In order to determine the motions of the shells after the collision, we impose 
the ``momentum conservation'' at the collision event; 
\begin{equation}
m_1u_1^a+m_2u_2^a=m_3 u_3^a+m_4u_4^a=:p^a
\,, \label{p-con}
\end{equation}
where $p^a$ is the conserved total 4-momentum of two shells
(see Appendix \ref{appendix_consistency}).
Using this conservation law 
(\ref{p-con}), in what follows, we will show how 
$u_3^a$ and $u_4^a$ are determined when $m_3$ and $m_4$ are fixed.
The 4-velocities $u_3^a$ and $u_4^a$ contain the information carried by two shells 
after collision, as we will show the details later. 

For this purpose, we write down $u_3^a$ in the linear combination form 
of $u_2^a$ and $n_2^a$,  and 
 describe the components of $u_3^a$ with respect to the coordinate basis in Region 3. 
This is because 
 the components of $u_2^a$ and $n_2^a$ with respect to 
the coordinate basis in Region 3 are given as the initial data before the collision. 
We also write down $u_4^a$ in the linear combination form of $u_1^a$ and $n_1^a$ 
and describe 
the components of $u_4^a$ with respect to the coordinate basis in Region 1
by the similar reason.

In general, scattering problems are extremely simplified in the center of mass frame. Hence, 
we define the dyad basis corresponding to the center of mass frame as
\begin{align}
u^a&=p^{-1}\left(m_1 u_1^a+m_2 u_2^a\right), \label{u}\\
n^a&=p^{-1}\left(m_1 n_1^a+m_2 n_2^a\right), \label{n}
\end{align}
where 
$$
p:=\sqrt{-p^a p_a}~.
$$

Here, we write the 4-velocities of Shell 3 and Shell 4 in the form
\begin{align}
u_3^a&=u^a\cosh\alpha+n^a\sinh\alpha, \label{u3}\\
u_4^a&=u^a\cosh\beta+n^a\sinh\beta. \label{u4}
\end{align}
The dyad components of the momentum conservation (\ref{p-con}) lead to
\begin{align}
m_3\cosh\alpha+m_4\cosh\beta&=p, \label{u-comp}\\
m_3\sinh\alpha+m_4\sinh\beta&=0. \label{n-comp}
\end{align}
From Eqs.~(\ref{u-comp}) and (\ref{n-comp}), we have
\begin{align}
m_4^2\cosh^2\beta&=p^2-2pm_3\cosh\alpha+m_3^2\cosh^2\beta. \label{u-comp-2} \\
m_4^2\sinh^2\beta&=m_3^2\sinh^2\alpha. \label{n-comp-2}
\end{align}
By subtracting each side of Eq.~(\ref{n-comp-2}) from that of Eq.~(\ref{u-comp-2}), we have 
$$
m_4^2=p^2-2pm_3\cosh\alpha+m_3^2,
$$
and hence
\begin{equation}
\cosh\alpha=\frac{p^2+m_3^2-m_4^2}{2pm_3}. \label{ch-alpha}
\end{equation}
By the similar manipulation, we also have
\begin{equation}
\cosh\beta=\frac{p^2+m_4^2-m_3^2}{2pm_4}.
\end{equation}
Since we consider the situation of Fig.~1, $\sinh\alpha$ is positive, whereas $\sinh\beta$ is negative;  
\begin{equation}
\sinh\alpha=+\sqrt{\cosh^2\alpha-1}~~~~{\rm and}~~~~\sinh\beta=-\sqrt{\cosh^2\beta-1}. \label{sh-beta}
\end{equation}

If we assume the proper masses $m_3$ and $m_4$ of the shells after the collision, 
$u_3^a$ and $u_4^a$ are determined by Eqs.~(\ref{u3}) and (\ref{u4}) with the coefficients 
given by Eqs.~(\ref{ch-alpha})--(\ref{sh-beta}). By using Eqs.~(\ref{u}) and (\ref{n}), 
we write down $u_3$ in the form of the linear combination of $u_J^a$ and $n_J^a$.

In order to write down the components of $u_3^a$,
we first replace $u_1^a$ and $n_1^a$ in $u^a$ and $n^a$ by the 
linear combinations of $u_2^a$ and $n_2^a$.  
For notational simplicity, we introduce 
\begin{align}
\Gm&:=-u_1^a u_{2a}=n_1^a n_{2a}, \label{Gamma-def} \\
V&:=u_1^a n_{2a}=-u_2^a n_{1a}.\label{V-def}
\end{align}
We have
\begin{align}
u_1^a&=\Gm u_2^a+V n_2^a, \label{trans-u1}\\
n_1^a&=V u_2^a+\Gm n_2^a. \label{trans-n1}
\end{align}
From the normalizations of $u_I^a$ and $n_I^a$ and the above equations, we have 
\begin{equation}
\Gm^2-V^2=1.
\end{equation} 
Since $u^a$ and $n^a$ are desicribed by the linear combinations of $u_2^a$ and $n_2^a$ 
by using Eqs.~(\ref{trans-u1}) and (\ref{trans-n1}), 
$u_3^a$ can also be described by the  
linear combinations of $u_2^a$ and $n_2^a$ through Eq.~(\ref{u3}). 
We also perform the similar manipulation for $u_4^a$ by using
\begin{align}
u_2^a&=\Gm u_1^a-V n_1^a, \label{trans-u2}\\
n_2^a&=-V u_1^a+\Gm n_1^a.\label{trans-n2}
\end{align}
As a result, we have
\begin{align}
u_3^a&=A_3u_2^a+B_3n_2^a, \label{u-3}\\
u_4^a&=A_4u_1^a+B_4n_1^a, \label{u-4}
\end{align}
where
\begin{align}
A_3&=\frac{1}{p}\left[\left(\Gm m_1+m_2\right)\cosh\alpha+V m_1\sinh\alpha\right],\\
B_3&=\frac{1}{p}\left[\left(\Gm m_1+m_2\right)\sinh\alpha+V m_1\cosh\alpha\right],\\
A_4&=\frac{1}{p}\left[\left(\Gm m_2+m_1\right)\cosh\beta -V m_2\sinh\beta\right],\\
B_4&=\frac{1}{p}\left[\left(\Gm m_2+m_1\right)\sinh\beta-V m_2\cosh\beta\right]. \label{B4}
\end{align}

\subsection{Components with respect to the coordinate basis}

From Eqs.~(\ref{u-3}) and (\ref{u-4}), we obtain the components of $u_3^a$ 
with respect to the coordinate basis in Region 3 and those of $u_4^a$ with respect to 
the coordinate basis in Region 1,
writing down the components of $u_2^a$ and $n_2^a$ 
 with respect to the coordinate basis in Region 3 as
\begin{align}
u_{2+}^\mu&=\left(\frac{e_{2+}}{f_3(r_2)},~ \dot{r}_2,~0,~0\right), \label{u2+}\\
n_{2+}^\mu&=\left(\frac{\dot{r}_2}{f_3(r_2)},~e_{2+},~0,~0\right), 
\label{u2+}
\end{align}
and the components of $u_1^a$ and $n_1^a$ 
with respect to the coordinate basis in Region 1 as
\begin{align}
u_{1-}^\mu&=\left(\frac{e_{1-}}{f_1(r_1)}, ~\dot{r}_1,~0,~0\right), \cr
n_{1-}^\mu&=\left(\frac{\dot{r}_1}{f_1(r_1)}, ~e_{1-},~0,~0\right)\,, 
\label{u1-}
\end{align}
where for notational simplicity, we introduce
\begin{align}
e_{1\pm}&:=\cE_1\mp\frac{m_1}{2r_1}, \\
e_{2\pm}&:=\cE_2\mp\frac{m_2}{2r_2}\,.
\end{align}

When we write $u_3^a$ and $u_4^a$ as
\begin{align}
u_{3+}^\mu&=\left(\frac{e_3}{f_3(r_3)},~ \dot{r}_3,~0,~0\right) \label{u3+}
\end{align}
and
\begin{align}
u_{4-}^\mu&=\left(\frac{e_4}{f_1(r_4)}, ~\dot{r}_4,~0,~0\right)
\,, \label{u4-}
\end{align}
where
\begin{align}
e_3&:=\cE_3-\frac{m_3}{2r_3}, \label{e3-def}\\
e_4&:=\cE_4+\frac{m_4}{2r_4}
\,, 
\end{align}
we find the relation between $e_{J\pm}$ and $e_K$.
Note that $e_I$, which correspond to the specific Killing energies 
for test 
particles, may describe the energies of the shells
but they are not conserved because of self-gravity effects of the shells. 

Hereafter, all components are evaluated at $r_1=r_2=r_3=r_4=\rc$, i.e., at the collision event. 
By using Eqs.~(\ref{u2+})--(\ref{u3+}), Eq.~(\ref{u-3}) leads to
\begin{align}
e_3&=A_3 e_{2+} +B_3\dot{r}_2,  \label{E3}\\
\dot{r}_3&= B_3e_{2+}+A_3\dot{r}_2. \label{p3}
\end{align}
By using Eqs.~(\ref{u2+})--(\ref{u3+}), Eq.~(\ref{u-4}) leads to
\begin{align}
e_4&=A_4 e_{1-} +B_4\dot{r}_1,  \label{E4}\\
\dot{r}_4&=B_4e_{1-}+A_4\dot{r}_1. \label{p4}
\end{align}

The components of $u_1^a$, $n_1^a$, $u_2^a$ and $n_2^a$ 
with respect to the coordinate basis in Region 2 are given by
\begin{align}
u_{2-}^\mu&=\left(\frac{e_{2-}}{f_2(\rc)},~ \dot{r}_2,~0,~0\right), \cr
n_{2-}^\mu&=\left(\frac{\dot{r}_2}{f_2(\rc)},~e_{2-},~0,~0\right), \cr
u_{1+}^\mu&=\left(\frac{e_{1+}}{f_2(\rc)}, ~\dot{r}_1,~0,~0\right), \cr
n_{1+}^\mu&=\left(\frac{\dot{r}_1}{f_2(\rc)}, ~e_{1+},~0,~0\right). \nonumber
\end{align}
Using these components, Eqs.~(\ref{Gamma-def}) and (\ref{V-def}) lead to  
\begin{align}
\Gm&=\frac{1}{f_2(\rc)}
\left(e_{1+}e_{2-}-\dot{r}_1\dot{r}_2\right), \label{Gamma}\\
V&=\frac{1}{f_2(\rc)}
\left(e_{2-}\dot{r}_1-e_{1+}\dot{r}_2\right). \label{V-sol}
\end{align}

Once we know the initial conditions of shells just before the collision 
($m_J$ and $u_J^a$ at the collision event) and the masses 
of shells just after the collision, 
$m_K$, we can obtain $\alpha$ and $\beta$ by Eqs.~(\ref{ch-alpha})--(\ref{sh-beta}), 
$\varGamma$ and $V$ by Eqs.~(\ref{Gamma}) 
and (\ref{V-sol}), and then $u_K^a$ by Eqs.~(\ref{E3})--(\ref{B4}); 
Note that the information about $u_J^a$ is equivalent to $e_{J\pm}$ and $\dot{r}_J$, whereas 
that about $u_K^a$ is equivalent to $e_K$ and $\dot{r}_K$.   
By the definition of $\cE_K$, the value of the metric function $f_4$ at the collision event is given by 
\begin{equation}
f_4(\rc)=f_3(\rc)+\frac{2m_3}{\rc}\left(e_3+\frac{m_3}{2\rc}\right)
=f_1(\rc)-\frac{2m_4}{\rc}\left(e_4-\frac{m_4}{2\rc}\right)
\label{metric-4}
\end{equation}
We will use Eq.~(\ref{metric-4}) for deriving the mass parameter $M_4$ of Region 4 in the next section.  

\section{Maximum energy extraction by the collision of charged shells}\label{sec:analytic}

Here we consider the situation in which the collision of two spherical shells 
occurs around a Reissner-Nordstr\"{o}m black hole. Each shell is assumed to be concentric to the black hole 
which is located in Region 1. Then, we study the maximum energy extraction 
from the black hole through the collisional Penrose process by two shells: 
Shell 4 falls into the black hole, whereas Shell 3 goes away to 
the infinity with the energy larger than the total energy carried 
initially by Shell 1 and Shell 2. In the test-shell limit $m_I/M_1\rightarrow0$, the present system  
recovers the situations studied in Refs.~\cite{Zaslavskii2012,Nemoto2013}.

The metric function of Region $I$ is given by 
$$
f_I(r)=1-\frac{2M_I}{r}+\frac{Q_I^2}{r^2},
$$
where $M_I$ and $Q_I$ are the mass  and charge parameters, respectively. 
The gauge one-form in the region $I$ is given by
$$
A_{I\alpha}=\left(-\frac{Q_I}{r},0,0,0\right).
$$

The charge of Shell $I$ is denoted by $q_I$. Gauss's law leads to 
$$
Q_2-Q_1=q_1,~~
Q_3-Q_2=q_2, ~~
Q_3-Q_4=q_3~~{\rm and}~~
Q_4-Q_1=q_4,
$$
or equivalently
$$
Q_2=Q_1+q_1,~~Q_3=Q_1+q_1+q_2~~{\rm and}~~Q_4=Q_1+q_1+q_2-q_3=Q_1+q_4.
$$
The above equations lead to the conservation of total charge through the collision: 
\begin{equation}
q_1+q_2=q_3+q_4. \label{q-cons}
\end{equation}

Equation (\ref{metric-4}) leads to
$$
M_4=M_3-\frac{Q_3^2-Q_4^2}{2\rc}-m_3\cE_3=M_1+\frac{Q_4^2-Q_1^2}{2\rc}+m_4\cE_4.
$$

In the case of the spherically symmetric system, almost all of the quasi-local energies 
proposed until now agree with the so-called Misner-Sharp energy~\cite{Misner1964}. 
In the present case, the Misner-Sharp energy 
within the sphere with the circumferential radius $r$ is given by
\begin{equation}
E_{\rm MS}(r)=M-\frac{Q^2}{2r}. \label{MS}
\end{equation}
Hence, the Misner-Sharp energy carried by Shell $I$ is given by 
$$
E_{\rm MS}(r_I)|_{+}-E_{\rm MS}(r_I)|_{-}=m_I\cE_I.
$$
If Shell $I$ has a non-vanishing charge, $m_I\cE_I$ depends on the radius $r_I$ due to 
the electric interaction. Then the energies of Shell 1, Shell 2 and Shell 3 found by the observers at infinity, 
are given by
\begin{align}
m_1E_1&=\lim_{r_1\rightarrow\infty}m_1\cE_1=M_2-M_1, \cr
m_2E_2&=\lim_{r_2\rightarrow\infty}m_2\cE_2=M_3-M_2, \cr
m_3E_3&=\lim_{r_3\rightarrow\infty}m_3\cE_3=M_3-M_4
=m_3\left(e_3|_{r=\rc}+\frac{m_3}{2\rc}\right)+\frac{q_3\bar{Q}_3}{\rc}, \label{E3-asym}
\end{align}
where 
$$
\bar{Q}_3:=\frac{Q_3+Q_4}{2}.
$$

Before proceeding to the non-linear analysis, it is intriguing to consider the case in which the test-shell 
approximation is applicable. 
 In this case, $\bar{Q}_3$ is regarded as the charge parameter of the fixed background spacetime, 
 whereas $q_3$ is the charge of Shell 3 going away to the infinity. 
Here, we assume $\bar{Q}_3>0$, and it should be noted that 
as long as the charge conservation (\ref{q-cons}) holds, Shell 3 can have arbitrary large charge $q_3$ 
fixing $\bar{Q}_3$ under  the test-shell approximation.  
Thus, if very large amount of charge is transferred from Shell 4 to Shell 3 by the collision  
so that $q_3\bar{Q}_3/\rc$ and then the extracted energy $m_3E_3$ can be much larger than 
the initial total energy of the shells $m_1E_1+m_2E_2$, 
the large amount of energy is extracted from the black-hole spacetime [see Eq.~(\ref{E3-asym})].  
There is no upper bound on the efficiency of the energy extraction, which is defined by 
\begin{equation}
\eta=\frac{m_3E_3}{m_1E_1+m_2E_2}. \label{eff-def}
\end{equation}
This is the case pointed out 
by Zaslavskii~\cite{Zaslavskii2012}. 
If we take into account the non-linearity of the shell contribution, it is not trivial whether there exist an upper bound on the efficiency $\eta$ or not, 
although the extracted energy is finite  since $q_3\bar{Q}_3=(Q_3^2-Q_4^2)/2\leq Q_3^2/2$ holds.

\subsection{Upper bound on the extracted energy}
\label{upper_bound}

Now we evaluate the upper bound on $m_3E_3$ by using the fact that 
the Misner-Sharp energy has a non-decreasing nature with respect to $r$, i.e.,  
in the direction of $n_I^a$ just on Shell $I$~\cite{Hayward1994}. 
From Eq.~(\ref{MS}), the following inequality should hold; 
\begin{equation}
M_1-\frac{Q_1^2}{2\rc}\leq M_4-\frac{Q_4^2}{2\rc} \leq M_3 -\frac{Q_3^2}{2\rc}. \label{non-decreasing}
\end{equation}
From Eq.~(\ref{non-decreasing}), we have
$$
m_3E_3=M_3-M_4<M_3-M_1+\frac{Q_1^2-Q_4^2}{2\rc}. 
$$
If the central black hole is extremal, i.e., $Q_1=M_1$, and $Q_4$ vanishes, we have 
$$
m_3E_3\leq M_3-M_1+\frac{M_1^2}{2\rc}. 
$$
Then the collision at the horizon radius $\rc=M_1$ in Region 1 gives the largest upper bound: 
\begin{equation}
m_3 E_3 <M_3-\frac{1}{2}M_1. \label{U-bound}
\end{equation}
This is  also easily understood from the view of the irreducible mass $M_{\rm ir}$ of the initial black hole. 
Since the initial black hole is described by an extremal Reissner-Nordstr\"{o}m  solution with the mass $M_1$, the irreducible mass 
is given by $M_{\rm ir}=M_1/2$. The rest energy $M_1-M_{\rm ir}=M_1/2$ is one from electromagnetic contribution and can be extracted by some mechanism.
As a result, The extracted energy $m_3 E_3$ is bounded by 
$M_3-M_{\rm ir}$, which is Eq. (\ref{U-bound}).

Inequality (\ref{U-bound}) leads to
\begin{equation}
M_4=M_3-m_3E_3>\frac{1}{2}M_1. \label{L-bound}
\end{equation}
Since Shell 4 will be absorbed into the black hole, the black hole  eventually becomes charge-neutral.  
Then, the area of its event horizon is larger than $4\pi M_1^2$ which is equal to the initial value of 
the extremal BH.  This result is consistent to the area law of the event horizon. 

It should be noted that the largest upper bound on $m_3E_3$ is achieved by the collision on the 
event horizon. This fact seems to imply that the BSW type collision is a necessary condition for the 
large efficiency $\eta$ in contrast to the test particle case. However, we will see in the following example 
that it is not necessarily the case.

\subsection{An example of almost maximum energy extraction}

In this subsection, we focus on the case of $m_3=m_1$ and $m_4=m_2$. 
By this restriction, the expressions of the energy-momentum transfer through the collision become 
so simple that we obtain analytically an example of almost maximum energy extraction.  
The same system has been studied by Ida and one of the present authors~\cite{Ida1999}, although 
they has not focused on the collisional Penrose process.  

We have
\begin{align}
\cosh\alpha&=\frac{m_1+m_2\varGamma}{p}, \label{ch-alpha-2}\\
\sinh\alpha&=\frac{m_2 V}{p}, \\
\cosh\beta&=\frac{m_1\varGamma+m_2}{p}, \\
\sinh\beta&=-\frac{m_1 V}{p}. \label{sh-beta-2}
\end{align}
Substituting Eqs.~(\ref{ch-alpha-2})--(\ref{sh-beta-2}) into Eqs.~(\ref{E3})--(\ref{p4}) 
and by using Eqs.~(\ref{f-1})--(\ref{f-4}), we have
\begin{align}
{\cal E}_3&={\cal E}_1-\frac{m_2}{\rc}\varGamma, \label{E-3}\\
\dot{r}_3&=\dot{r}_1-\frac{m_2}{\rc}V, \\
{\cal E}_4&={\cal E}_2+\frac{m_1}{\rc}\varGamma, \\
\dot{r}_4&=\dot{r}_2-\frac{m_1}{\rc}V. \label{r-4}
\end{align}
It is worthwhile to notice that $\dot{r}_3<\dot{r}_1$ and $\dot{r}_4<\dot{r}_2$ hold because of $V>0$. 
Note also that $\varGamma>0$ holds by its definition, and hence ${\cal E}_3<{\cal E}_1$ 
and ${\cal E}_4>{\cal E}_2$ hold.    


We again assume that the black hole is initially extremal and finally charge-neutral 
as a result of the absorption of Shell 4 by the black hole: 
\begin{eqnarray}
Q_1=M_1\quad{\rm and}\quad Q_4=0. \label{Q-assume}
\end{eqnarray}
Furthermore, we assume 
\begin{equation}
q_1=0 . \label{q1-assume}
\end{equation}
Since we assume the collision takes place near the horizon, 
we write the circumferential radius at the collision event, $r=\rc$, in the form of 
\begin{equation}
\rc=\frac{M_1}{1-\varepsilon}, \label{r-collision}
\end{equation}
with $0<\varepsilon \ll1$.  

Hereafter, a character with a tilde denotes a quantity normalized by the initial mass of the black hole, $M_1$,    
i.e., $\tq_I \equiv q_I/M_1$ and $\tm_I\equiv m_I/M_1$.
Since Shell 1 and Shell 2 approach the black hole from infinity,  
$E_J$ should be larger than or equal to unity. We focus on the case that $E_J$ is of order unity. 

Together with Eqs.~(\ref{Q-assume})--(\ref{r-collision}), Eq.~(\ref{E-3}) leads to  
\begin{eqnarray}
 E_3 &=& E_1 +\frac{1 +2\tq_2 +\tq_2^2}{2\tm_1}\left(1-\veps\right) -\tm_2\varGamma\left(1-\veps\right).
 \label{E3-exact}
\end{eqnarray}
The assumptions (\ref{Q-assume})--(\ref{r-collision}) lead to 
\begin{align}
 f_2(\rc) &= \veps^2 -2\tm_1E_1\left(1-\veps\right),  \label{f2} \\
 f_3(\rc)&= \veps^2-2(\tm_1E_1+\tm_2E_2)(1-\veps)+\tq_2(2+\tq_2)(1-\veps)^2. \label{f3}
\end{align}
Since the collision should occur outside the black hole, we have from Eqs.~(\ref{f2}) and (\ref{f3}) the 
following constraints
\begin{align}
\frac{\veps^2}{2(1-\veps)}&>\tm_1 E_1 . \label{UB-E1}\\
\frac{\veps^2}{2(1-\veps)}&>\tm_1E_1+\tm_2E_2-\tq_2\left(1+\frac{\tq_2}{2}\right)(1-\veps). \label{UB-E2}
\end{align}
Equation (\ref{UB-E1}) implies that  
$\tm_1$ should be at most of order $\veps^2$, and then Eq.~(\ref{UB-E2}) implies that 
both of $\tm_2$ and $\tq_2$ should also be at most of order $\veps^2$.  Hereafter we assume
$$
\tm_1,~~\tm_2,~~\tq_2={\cal O}(\veps^2)~~~{\rm and}~~~\frac{\tq_2}{\tm_1},~~\frac{\tq_2}{\tm_2},
~~\frac{\tq_2}{\veps^2}={\cal O}(\veps^0).
$$

We consider the situation in which $\dot{r}_1\dot{r}_2>0$ holds at the collision event. 
Then, $\varGamma$ is approximately estimated at
\begin{align}
 \varGamma &\sim \frac{1}{2}\left( \frac{e_{1+}}{e_{2-}} +\frac{e_{2-}}{e_{1+}} \right) \cr
&\sim \frac{\left(\tm_2E_1\right)^2+\left(\tm_2E_2-\tq_2\right)^2}{2\tm_2E_1\left(\tm_2E_2-\tq_2\right)}
-\frac{\left(\tm_2E_1\right)^2-\left(\tm_2E_2-\tq_2\right)^2}{2\tm_2E_1\left(\tm_2E_2-\tq_2\right)^2}
\tq_2\veps
+{\cal O}(\veps^2).
\end{align}
Therefore, the asymptotic energy of Shell 3 is given by
\begin{equation}
 E_3=\frac{1}{2\tm_1}-\frac{\veps}{2\tm_1}+E_1+\frac{\tq_2}{\tm_1}-\frac{\tq_2}{\tm_1}\veps+{\cal O}(\veps^2)\,,\label{cE3}
\end{equation}
which gives  the energy extracted from the system explicitly as 
\begin{equation}
 m_1 E_3 = \frac{M_1}{2}\left[1-\veps+2\left(\tm_1E_1+\tq_2\right)+{\cal O}(\veps^3)\right]. \label{mE3}
\end{equation}
As we discussed in \S \ref{upper_bound}, the upperbound is given by 
Eq. (\ref{U-bound}). If the asymptotic specific energies $E_1$ and $E_2$ are not so large, 
i.e., $M_3\sim M_1$, the above energy extraction (\ref{mE3}) gives almost maximal value if $\varepsilon \ll 1$ (the 
 collision occurs near the horizon).
The present result implies that the collisional Penrose process of two 
charged shells with very small masses can achieve the almost maximum energy extraction, if the black hole becomes 
finally charge-neutral.

As for the efficiency of the energy extraction $\eta$, we have : 
\begin{equation}
 \eta \equiv \frac{m_1E_3}{m_1E_1 +m_2E_2}
= \frac{\tm_1E_3}{\tm_1E_1 +\tm_2E_2}={\cal O}(\veps^{-2}). \label{eff-eg}
\end{equation}
Equation (\ref{eff-eg}) implies that there is no upper bound on $\eta$. 
This is because 
the maximally extracted energy $m_3 E_3\sim M_1/2$ is finite even if the input initial energy of two shells, $m_1E_1 +m_2E_2$, 
is infinitely small.

We also have
\begin{equation}
M_4=M_1+m_1E_1+m_2E_2-m_3E_3=\frac{M_1}{2}\left[1+\veps+{\cal O}(\veps^2)\right]\,, \label{M4-min}
\end{equation}
which guarantees the consistency with the area law of the black hole
because $A_4=4\pi (2M_4)^2>A_1=4\pi M_1^2$, as mentioned below Eq.~(\ref{L-bound}). 

In order to extract the energy from a black hole, in addition to the above energy argument, we must impose one additional condition, which is that Shell 3 must move outward to infinity.
However, by the reason mentioned below Eq.~(\ref{r-4}), $\dot{r}_1<0$ implies $\dot{r}_3<0$ just after the collision. 
Hence Shell 3 has to bounce off the potential barrier so that it goes away to infinity. As shown below, 
this bounce will happen under some possible condition. 

From Eq.~(\ref{eq:motion-K}), the energy equation of Shell 3 is written in the form,
$$
\dot{r}_3^2 +V(r_3)=0,
$$
where the effective potential $V(r)$ is given by
\begin{equation}
 V(r)=f_3(r)+w(r).\label{Veff}
\end{equation}
with  introducing a function $w(r)$ defined by
$$
w(r):=-e_3^2=-\left(E_3-\frac{M_1\left[\left(1+\tq_2\right)^2+\tm_1^2\right]}{2\tm_1 r}\right)^2
\,.
$$

The function $w(r)$ has a zero point and a maximum at the identical circumferential radius 
$r=r_{\rm m}$ in the domain of $r>0$.
By using Eq.~(\ref{mE3}), we find 
\begin{equation}
r_{\rm m}=\frac{M_1\left[\left(1+\tq_2\right)^2+\tm_1^2\right]}{2\tm_1E_3}
=\frac{M_1}{1-\veps}\times\left[1-2\tm_1E_1+{\cal O}(\veps^3)\right]<\rc, \label{rm-rc}
\end{equation}
or the explicit form up to the second order of $\veps$ as
\begin{equation}
r_{\rm m}=M_1\left[1+\veps+\veps^2-2\tm_1E_1+{\cal O}(\veps^3)\right]. \label{rm-def}
\end{equation}
The metric function $f_3(r)$ is rewritten in the form 
$$
f_3(r)=1-\frac{2M_1}{r}\left(1+\tm_1E_1+\tm_2E_2\right)+\frac{M_1^2}{r^2}\left(1+\tq_2\right)^2.
$$
It is easy to see that $f_3(r)$ is a monotonically increasing function in the domain $r>M_1$. The larger root 
of $f_3=0$ corresponds to the horizon radius $r_{\rm h}$ in Region 3;
\begin{align}
r_{\rm h}&=M_1\left[1+\tm_1E_1+\tm_2E_2+\sqrt{\left(1+\tm_1E_1+\tm_2E_2\right)^2-\left(1+\tq_2\right)^2}\right] \cr
&=M_1\left[1+\sqrt{2\left(\tm_1E_1+\tm_2E_2-\tq_2\right)}+\tm_1E_1+\tm_2E_2+{\cal O}(\veps^3)\right]. 
\label{rh-def}
\end{align}
From Eq.~(\ref{UB-E2}), we have
$$
\sqrt{2(\tm_1E_1+\tm_2E_2-\tq_2)}<\veps+\frac{\veps^2}{2}-\tq_2+{\cal O}(\veps^3),
$$
which leads to, together with Eqs.~(\ref{rm-def}) and (\ref{rh-def}), 
\begin{align}
r_{\rm m}-r_{\rm h}&=M_1\left[\veps+\veps^2-3\tm_1E_1-\tm_2E_2-\sqrt{2(\tm_1E_1+\tm_2E_2-\tq_2)}
+{\cal O}(\veps^3)\right] \cr
&>\frac{\veps^2}{2}-3\tm_1E_1-\tm_2E_2+\tq_2+{\cal O}(\veps^3)
\,, \label{rm-rh}
\end{align}
as shown the details in Appendix \ref{appendix_additional_condition}. 

In order  for Shell 3 to bounce off before the horizon, we impose a sufficient condition, 
which is  $r_{\rm m}>r_{\rm h}$, which gives one additional condition for $\tm_1E_1$ and $\tm_2E_2$  such that 
\begin{equation}
\frac{\veps^2}{2}-3\tm_1E_1-\tm_2E_2+\tq_2+{\cal O}(\veps^3)>0.\label{additional}
\end{equation}
With this  condition,  
we find
\begin{equation}
r_{\rm h}<r_{\rm m}<\rc ~~~~~~{\rm for}~~\tm_1E_1+\tm_2E_2-\tq_2\geq0, \label{rh-rm-rc}
\end{equation}
and
$$
r_{\rm m}<\rc~~{\rm and}~~f_3(r)>0~~~~~~{\rm for}~~\tm_1E_1+\tm_2E_2-\tq_2<0.
$$
 Since $w(r_{\rm m})=0$ and $f_3(r_{\rm m})>0$ hold, we have $V(r_{\rm m})>0$. 
 It is not so difficult to obtain $V(\rc)=-E_1^2+{\cal O}(\veps^2)<0$. 
 Together with the inequality (\ref{rh-rm-rc}), 
 these facts imply that Shell 3 initially moving to the black hole should bounce off at 
 the potential barrier with the circumferential radius 
 $r_{\rm b}$ which satisfies $r_{\rm m}<r_{\rm b}<\rc$, and then go away to the infinity. 
 Shell~3 can carry the huge energy extracted from the black hole to the infinity. 
 
We should recognize that the condition (\ref{additional}) gives a constraint 
on the initial energies of two shells (and charge of Shell 2).
If we wish to get the large efficiency, the initial energies must be small.
This is because the maximum extracted energy is finite and fixed, and 
the efficiency becomes large if the initial energies are small. 
 Note that the inequality (\ref{additional}) is consistent with Eqs.~(\ref{UB-E1}) and (\ref{UB-E2}).

\begin{figure}[!h]
\centering\includegraphics[width=2.5in]{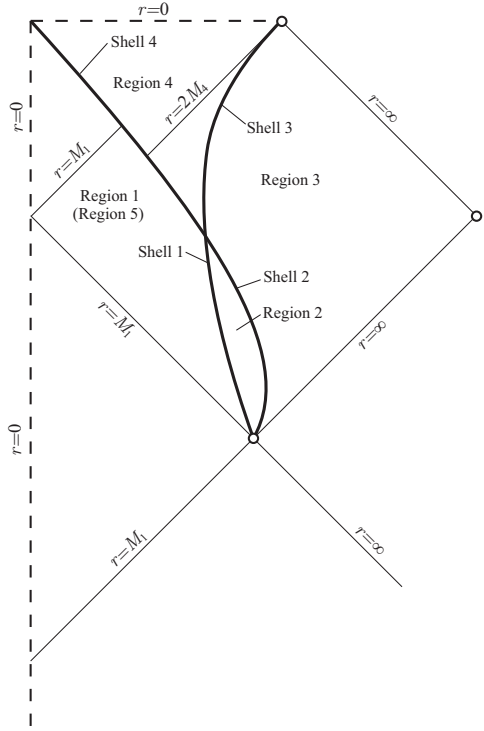}
\caption{The Penrose diagram of the spacetime in which the two shells collide with each other around 
an extremal Reissner-Nordstr\"{o}m black hole. Since all of the charges are carried by Shell 3, 
the final state of the spacetime is given by a neutral Schwarzschild black hole. }
\label{Penrose}
\end{figure}

 In Fig.~\ref{Penrose}, we depict 
 the Penrose diagram of the spacetime in which the collision described in this subsection occurs. 
 Two massive charged shells are initially falling toward 
 an extremal Reissner-Nordstr\"{o}m black hole, and then collide near the horizon.
 After a large amount of charge transfer at the collision, 
  Shell 3 will bounce off at the potential barrier and then goes away to infinity with a huge amount of energy.
 The finial spacetime turns to be a neutral Schwarzschild black hole.

Here, we should note that the BSW collision does not occur in the present 
situation, where the BSW collision means 
that the center-of-mass energy at the collision near the horizon becomes unboundedly large, i.e., 
$p$ diverges in the limit of $\varepsilon \rightarrow 0$.
As we show in Appendix \ref{appendix_BSW}, $p$ does not diverge as $\varepsilon \rightarrow 0$ in the present example,
although we find the almost maximum energy extraction from the black hole.

\section{Conclusion}\label{sec:conclusion}

We have investigated the energy extraction process from a Reissner-Nordstr\"{o}m black hole through 
the collision of two spherical charged shells by taking into account the self-gravity of the shells. 
We have derived the conditions for the shells just after the collision by imposing the conservation of 
total 4-momentum, where the mass and charge transfers between the shells are allowed. 
Then, from the monotonicity of the Misner-Sharp mass, we show that the extracted energy is bounded 
from above, and the upper bound is the half of the ADM energy of the
initial black hole, which is consistent with the area law of the black hole. Furthermore, from this consideration, 
we find the following conditions for the large energy extraction;  
\begin{enumerate}
\item  The collision event must be very close to the event horizon;
\item The initial black hole is nearly the extremal Reissner-Nordstr\"{o}m one; 
\item The final spacetime is a charge-neutral Schwarzschild black hole.
\end{enumerate}
Finally, we have shown one scenario of almost maximum energy extraction, 
in which the BSW collision does not take place.

As for the BSW collision, as we have shown one example in Appendix C, 
we expect that it will not lead to the maximum energy extraction.
The collision with the infinite center-of-mass energy may take place inside the horizon radius 
which is necessarily larger than the initial horizon radius $r=M_1$. 
In order to extract the energy, the collision point must be outside the horizon. 
Here we should again note that the collision event at $r=M_1$ will be a necessary condition 
for the maximum energy extraction. Hence the BSW collision will not achieve it. 
However, if a BSW-{\it like} collision is possible, which means 
the center-of-mass energy is not infinite but very large, so that large
energy extraction is possible, we may find new particle with large mass through 
a collision near a black hole horizon similar to the process found by Nemoto et al by invoking the test 
particle approximation~\cite{Nemoto2013}, and reveal new aspect of high energy physics.
The work on a BSW-like collision in the present model with two charged shells
and the possible energy extraction
will be published elsewhere.

Since an extremely charged black hole may not exist in nature, 
it is more interesting to study a rapidly rotating black hole with  collisional spinning particles\cite{spinning1, spinning2, spinning3, spinning4}. If we can extract the maximum energy determined by the irreducible mass, 
i.e., 
$$
M-M_{\rm ir}=M-{M\over \sqrt{2}}
\,,
$$
by the collisional Penrose process, 
we will find most effective energy extraction method from a rotating black hole. This study is also in progress. 

\section*{Acknowledgments}

This work was supported in part by JSPS KAKENHI Grant Numbers JP25400265 (KN), JP16K05362 (KM), 
17H06357 (KM) and JP17H06359 (KM).

\appendix

\section{Consistency check of momentum conservation}
\label{appendix_consistency}
It should be noted that, by the definition of 
$\cE_J$ and $\cE_K$, i.e., Eqs.~(\ref{eq:energy-J}) and (\ref{eq:energy-K}), 
the following relation is trivially satisfied at the collision event: 
\begin{equation}
m_1\cE_1+m_2 \cE_2=m_3 \cE_3+m_4\cE_4. \label{E-con}
\end{equation}
Since it seems to be non-trivial whether the relation (\ref{E-con}) is consistent 
with the momentum conservation~(\ref{p-con}),
which is our ansatz, 
we will show in this appendix that it is the case. 

It is impossible to directly derive Eq.~(\ref{E-con}) from Eqs.~(\ref{E3}), (\ref{p3}), (\ref{E4}) and (\ref{p4}) 
obtained from the momentum conservation 
(\ref{p-con}), and hence we rewrite them in the appropriate form for our purpose. 
There are several useful relations derived by using Eq.~(\ref{eq:motion-J});
\begin{align}
\Gm\dot{r}_1&=\dot{r}_2+Ve_{1+}, \label{f-1}\\
\Gm\dot{r}_2&=\dot{r}_1-Ve_{2-}, \\
V\dot{r}_1&=\Gm e_{1+}-e_{2-},\\
V\dot{r}_2&=e_{1+}-\Gm e_{2-}. \label{f-4}
\end{align}
By using the above relations, Eqs. (\ref{E3}), (\ref{p3}), (\ref{E4}) and (\ref{p4}) are rewritten in the form
\begin{align}
e_3&=\frac{1}{p}\biggl[\left(m_1e_{1+}+m_2e_{2+}-\frac{m_1m_2}{r}\Gm\right)\cosh\alpha
+\left(m_1\dot{r}_1+m_2\dot{r}_2-\frac{m_1m_2}{r}V\right)\sinh\alpha\biggr], \label{ep3}\\
\dot{r}_3&=\frac{1}{p}\biggl[\left(m_1e_{1+}+m_2e_{2+}-\frac{m_1m_2}{r}\Gm\right)\sinh\alpha
+\left(m_1\dot{r}_1+m_2\dot{r}_2-\frac{m_1m_2}{r}V\right)\cosh\alpha\biggr], \\
e_4&=\frac{1}{p}\biggl[\left(m_1e_{1-}+m_2e_{2-}+\frac{m_1m_2}{r}\Gm\right)\cosh\beta
+\left(m_1\dot{r}_1+m_2\dot{r}_2-\frac{m_1m_2}{r}V\right)\sinh\beta\biggr], \label{ep4}\\
\dot{r}_4&=\frac{1}{p}\biggl[\left(m_1e_{1-}+m_2e_{2-}+\frac{m_1m_2}{r}\Gm\right)\sinh\beta
+\left(m_1\dot{r}_1+m_2\dot{r}_2-\frac{m_1m_2}{r}V\right)\cosh\beta\biggr]. \label{r4-dot}
\end{align}
Then, by using Eqs.~(\ref{u-comp}), (\ref{n-comp}) and (\ref{ch-alpha}), Eqs.~(\ref{ep3}) and (\ref{ep4}) lead to
Eq.~(\ref{E-con}). 

\section{Additional sufficient condition for the bounce of Shell 3}
\label{appendix_additional_condition}
To make the analysis simple, we introduce the following three parameters,
\begin{align}
 \tm_1 = \mu_1\veps^2, \tm_2 = \mu_2\veps^2, \tq_2 =-\delta_2\veps^2,
\end{align}
satisfying $0<\mu_1<1,\ 0<\mu_2<1$ and $-1<\delta_2<1$.

\begin{align}
 r_{\rm h}&=M_1\left[1+\tm_1E_1+\tm_2E_2+\sqrt{\left(1+\tm_1E_1+\tm_2E_2\right)^2-\left(1+\tq_2\right)^2}\right] \cr
 &=M_1\left[1+\veps\sqrt{2\left(\mu_1E_1+\mu_2E_2+\delta_2\right)}+\left(\mu_1E_1+\mu_2E_2\right)\veps^2 +{\cal O}(\veps^3)\right]. 
\label{rh}
\end{align}
Eq.~(\ref{UB-E2}) under the condition $\veps\ll 1$ yields
\begin{eqnarray}
 \frac{\veps^2}{1-\veps} < 2\left(\mu_1E_1 +\mu_2E_2 +\delta_2\right)\veps^2
  -2\delta_2\veps^3 -\delta_2^2\veps^4 +\mathcal{O}(\veps^5),\nonumber\\
 \frac{1}{1-\veps} +2\delta_2\veps +\delta_2^2\veps^2 +\mathcal{O}(\veps^3) < 2\left(\mu_1E_1 +\mu_2E_2 +\delta_2\right),\nonumber\\
 \sqrt{\frac{1}{1-\veps} +2\delta_2\veps +\delta_2^2\veps^2 +\mathcal{O}(\veps^3)} < \sqrt{2\left(\mu_1E_1 +\mu_2E_2 +\delta_2\right)},\nonumber\\
 1+\frac{1}{2}\left(1+2\delta_2\right)\veps +\mathcal{O}(\veps^2) < \sqrt{2\left(\mu_1E_1 +\mu_2E_2 +\delta_2\right)}.
\end{eqnarray}
which leads to, together with Eqs.~(\ref{rm-def}) and (\ref{rh}), 
\begin{align}
r_{\rm m}-r_{\rm h}&=M_1\left[\veps +\veps^2 -3\mu_1E_1\veps^2 -\mu_2E_2\veps^2 -\veps\sqrt{2(\tm_1E_1+\tm_2E_2+\tq_2)}
+{\cal O}(\veps^3)\right] \cr
&>\frac{\veps^2}{2} -3\mu_1E_1\veps^2 -\mu_2E_2\veps^2 -\delta_2\veps^2 +{\cal O}(\veps^3). \label{rm-rh-2}
\end{align}
Therefore, the following relation among parameters becomes a sufficient
condition for Shell 3 to bounce off:
\begin{align}
 \frac{1}{2} > 3\mu_1E_1 +\mu_2E_2 +\delta_2\,,\label{condition_bounce}
\end{align}
which gives one additional condition (\ref{additional}) imposed in the text.

\begin{figure}[!h]
\centering\includegraphics[width=5.in]{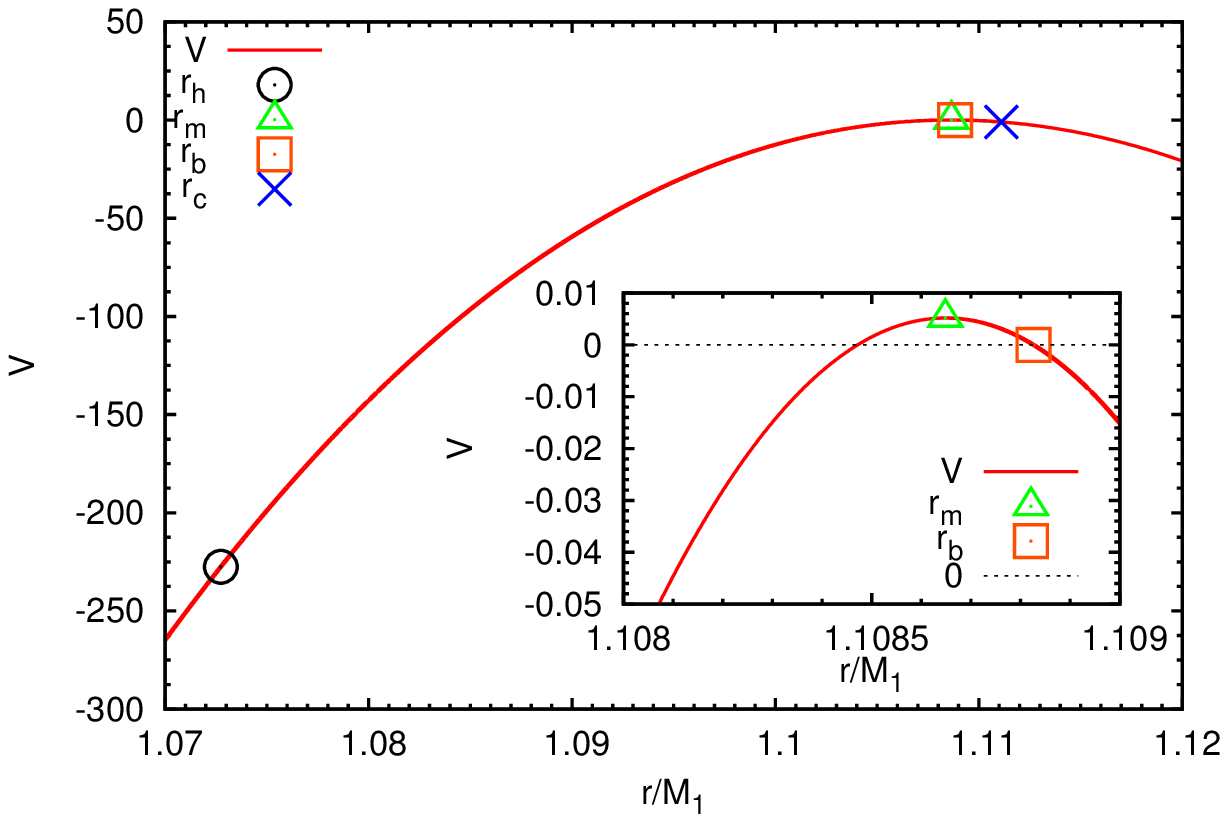}
\caption{An example of effective potential~\eqref{Veff} for Shell 3 with a given
 parameter set which achives a large energy extraction.
 We choose $\veps=0.1$, $\mu_1=\mu_2=0.1$, $E_1=E_2=1$
 and $\delta_2=0.05$ so that Eq.~\eqref{condition_bounce} is satisfied.
 }
\label{Bounce}
\end{figure}


\section{BSW collision v.s. non-BSW collision}

\label{appendix_BSW}

The BSW collision is defined as that with the extremely large collision energy in the 
center of mass frame. The collision energy in the center of mass is equal to $p$ which is written in the form
$$
p=\sqrt{m_1^2+m_2^2+2m_1m_2\varGamma}.
$$
The above equation implies that the large $p$ is equivalent to the large $\varGamma$. 
The normalization condition of the 4-velocities of Shell 1 and Shell 2 lead to
\begin{align}
e_{1+}&=\sqrt{\dot{r}_1^2+f_2(r_{\rm c})}, \cr
e_{2-}&=\sqrt{\dot{r}_2^2+f_2(r_{\rm c})}.\nonumber
\end{align}
Hence, from Eq.~(\ref{Gamma}), we have
\begin{align}
 \varGamma=\frac{\sqrt{(\dot{r}_1^2+f_2)(\dot{r}_2^2+f_2)}
 -\dot{r}_1\dot{r}_2}{f_2}
 \label{C1}
\end{align}
We consider the collision near the horizon in Region 2, i.e., $0<f_2\ll1$. 
Thus, we write the radius at the collision event in the form
\begin{equation}
\rc=\frac{M_2+\sqrt{M_2^2-Q_2^2}}{1-\epsilon},
\end{equation}
where we have assumed $|Q_2|\leq M_2$: in the limit of $\epsilon\rightarrow0$, the collision 
occurs at the horizon in Region 2.  
We also assume $\dot{r}_1\dot{r}_2 >0$.
Since
Eq. (\ref{C1})  is rewritten as
\begin{align}\varGamma=\frac{\dot{r}_1^2+\dot{r}_2^2+f_2}{\sqrt{(\dot{r}_1^2+f_2)(\dot{r}_2^2+f_2)}+\dot{r}_1\dot{r}_2}
\simeq\frac{\dot{r}_1^2+\dot{r}_2^2}{2\dot{r}_1\dot{r}_2}
\,,
\end{align}
near the horizon ($f_2\simeq 0$),
the BSW collision implies that 
either $\dot{r}_1$ or $\dot{r}_2$ vanishes   
as $\epsilon \rightarrow 0$.

Since $|\dot{r}_1|\approx e_{1+}$ and $|\dot{r}_2|\approx e_{2-}$ near the horizon, we have to evaluate $e_{1+}$ and $ e_{2-}$ 
at the collision point.
From the definition, 
we find 
\begin{align*}
e_{1+}(r_c)&={\cal E}_1-{m_1\over 2r_c}={r_c\over 2m_1}\left[f_1(r_c)-f_2(r_c)\right]-{m_1\over 2r_c}=E_1+{Q_1^2-Q_2^2\over 2m_1 r_c}-{m_1\over 2r_c}
\,,
\\
e_{2-}(r_c)&={\cal E}_2+{m_2\over 2r_c}={r_c\over 2m_2}\left[f_2(r_c)-f_3(r_c)\right]+{m_2\over 2r_c}=E_2+{Q_2^2-Q_3^2\over 2m_2 r_c}+{m_2\over 2r_c}
\,.
\end{align*}

In our example, we can easily show that 
both $|\dot{r}_1|$ and $|\dot{r}_2|$ are finite as follows:\\Using our ansatz, $Q_1=M_1$ and 
$q_1=0$, 
we find
\begin{align*}
e_{1+}(r_c)&\approx E_1+O(\epsilon)
\,,
\\
e_{2-}(r_c)&\approx E_2-{q_2 M_1\over m_2 r_c}+O(\epsilon)
\,,
\end{align*}
which yields $|\dot{r}_1| $ and $|\dot{r}_2|
$ are finite. As a result,  $\varGamma$ is also finite, and then 
$p$ does not diverge near the horizon. It is not the BSW collision.

When we find the BSW collision ?
One of $\dot{r}_1^2$ or $\dot{r}_2^2$ must vanish near the horizon. 
Then we consider the case that $\dot{r}_1^2 \simeq \alpha^2
 f_2$ with $\alpha>0$ whereas $\dot{r}_2^2$ is finite. 
This can be realized if we assume 
\begin{equation}
|Q_1|<M_1 ~~~{\rm and}~~~|Q_2|=M_2. \label{BSW1}
\end{equation} 
Since 
\begin{equation}
e_{1+}=E_1\left[1-\frac{M_2^2-Q_1^2+m_1^2}{2(M_2-M_1)r_c}\right].
\end{equation}
we obtain $e_{1+}=E_1\sqrt{f_2} $  
 if and only if 
$$
\frac{M_2^2-Q_1^2+m_1^2}{2(M_2-M_1)}=M_2
$$
is satisfied, where we have $f_2=(1-M_2/r)^2$.

The root of the above equation, which satisfies $M_2>M_1$, is 
\begin{equation}
M_2=M_1+\sqrt{M_1^2-Q_1^2+m_1^2}. \label{BSW2}
\end{equation}
From the normalization condition of the 4-velocity of Shell 1, i.e., $\dot{r}_1^2=e_{1+}^2-f_2$, we find  
$$
\dot{r}_1^2=\alpha^2 f_2,
$$
where 
$$
\alpha^2 =E_1^2-1
$$
with $$
E_1:=\frac{M_2-M_1}{m_1}=\sqrt{\frac{M_1^2-Q_1^2}{m_1^2}+1}~>1.
$$

As for $\dot{r}_2^2$, we obtain  $|\dot{r}_2|\approx e_{2-}\propto \sqrt{f_2}$, if and only if 
$$
M_3^2-Q_3^2=m_2^2(E_2^2-1)\sim O(\epsilon^2)
\,.
$$
Hence if we assume that Region 3 spacetime is not extreme, i.e., $M_3^2-Q_3^2\sim O(\epsilon^0)$, 
$\dot{r}_2^2$ is finite near the horizon.

We then find
\begin{align}
\varGamma\simeq
\frac{\sqrt{1+\alpha^2}-\alpha}{\sqrt{f_2}}~|\dot{r}_2|.
\end{align}
and  $p$ will diverge near the horizon, which corresponds to the BSW collision.  
Hence 
the BSW collision between 
Shell 1 and Shell 2 is possible
in the case that Eqs.~(\ref{BSW1}) and (\ref{BSW2}) are satisfied and Region 3 spacetime is not extreme. 
The horizon radius of Region 3 is larger than that of Region 2:
$$
M_3+\sqrt{M_3^2-Q_3^2}=M_2+m_2E_2+\sqrt{M_3^2-Q_3^2}>M_2.
$$
This fact implies that the present  
BSW collision necessarily occurs inside a black hole. 
In order to extract energy, the collision point must be outside 
the horizon. We then expect that the BSW collision
may not lead to the maximum energy extraction.


\begin{thebibliography}{9}

\bibitem{Penrose1969}
R.~Penrose,
Riv. Nuovo Cim. {\bf 1}, 252 (1969), [Gen. Rel. Grav.34,1141(2002)].

\bibitem{Bardeen1972}
J.~M. Bardeen, W.~H. Press and S.~A. Teukolsky,
Astrophys. J. {\bf 178}, 347 (1972).

\bibitem{Wald1974}
R.~M. Wald,
Astrophys. J. {\bf 191}, 231 (1974).

\bibitem{Piran1975}
T.~{Piran}, J.~{Shaham} and J.~{Katz},
Astrophys. J. Lett. {\bf 196}, L107 (1975).

\bibitem{Banados2009}
M.~Banados, J.~Silk and S.~M. West,
Phys. Rev. Lett. {\bf 103}, 111102 (2009), [0909.0169].

\bibitem{Bejger2012}
M.~Bejger, T.~Piran, M.~Abramowicz and F.~Hakanson,
Phys. Rev. Lett. {\bf 109}, 121101 (2012), [1205.4350].

\bibitem{Harada2012}
T.~Harada, H.~Nemoto and U.~Miyamoto,
Phys. Rev. D {\bf 86}, 024027 (2012), [1205.7088],
[Erratum: Phys. Rev.D86,069902(2012)].

\bibitem{Schnittman2014}
J.~D. Schnittman,
Phys. Rev. Lett. {\bf 113}, 261102 (2014), [1410.6446].

\bibitem{Ogasawara2015}
K.~Ogasawara, T.~Harada and U.~Miyamoto,
Phys. Rev. D {\bf 93}, 044054 (2016), [1511.00110].

\bibitem{Leiderschneider2016}
E.~{Leiderschneider} and T.~{Piran},
Phys. Rev. D {\bf 93}, 043015 (2016), [1510.06764].

\bibitem{Zaslavskii2016}
O.~B. Zaslavskii,
Phys. Rev. D {\bf 94}, 064048 (2016), [1607.00651].

\bibitem{Hod2016}
S.~Hod,
Phys. Lett. {\bf B759}, 593 (2016), [1609.06717].

\bibitem{Berti2014}
E.~Berti, R.~Brito and V.~Cardoso,
Phys. Rev. Lett. {\bf 114}, 251103 (2015), [1410.8534].

\bibitem{Patil2015}
M.~Patil, T.~Harada, K.-i. Nakao, P.~S. Joshi and M.~Kimura,
Phys. Rev. D {\bf 93}, 104015 (2016), [1510.08205].

\bibitem{Zaslavskii2010}
O.~B. Zaslavskii,
JETP Lett. {\bf 92}, 571 (2010), [1007.4598]

\bibitem{Zaslavskii2012}
O.~B. Zaslavskii,
Phys. Rev. D {\bf 86}, 124039 (2012), [1207.5209].

\bibitem{Nemoto2013}
H.~{Nemoto}, U.~{Miyamoto}, T.~{Harada} and T.~{Kokubu},
Phys. Rev. D {\bf 87}, 127502 (2013), [1212.6701].

\bibitem{Kimura2011}
M.~{Kimura}, K.~{Nakao} and H.~{Tagoshi},
Phys. Rev. D {\bf 83}, 044013 (2011), [1010.5438]

\bibitem{Wald}
R.M.~{Wald}, 
{\it General Relativity} (The University of Chicago Press, Chicago, 1984). 


\bibitem{Israel}
W. Israel, Nuovo Cim. {\bf 44B}, 1 (1966). 

\bibitem{Misner1964}
C.W.~Misner and D.H.~Sharp,
Phys. Rev. {\bf 136}, B571 (1964).

\bibitem{Hayward1994}
S.~A. Hayward,
Phys. Rev. D {\bf 53}, 1938 (1996), [gr-qc/9408002].

\bibitem{Ida1999}
D.~Ida and K.-i~Nakao,
Prog. Theor. Phys. {\bf 101}, 989 (1999).

\bibitem{spinning1}
C. Armaza, M. Banados, and B. Koch, Class. Quantum Grav. {\bf 33}, 105014 (2016).
\bibitem{spinning2}
O. B. Zaslavskii, Europhys. Lett. {\bf 114}, 30003  (2016) 
\bibitem{spinning3}
M.-Y. Guo and S.-J. Gao, Phys. Rev. {\bf D93}, 084025 (2016).
\bibitem{spinning4}
Y.-P. Zhang, B.-M. Gu, S.-W. Wei, J. Yang and Y.-X. Liu, Phys. Rev. {\bf D94}, 124017 (2016).

\end{thebibliography}
%

\end{document}